\documentclass[a4paper]{jpconf}
\usepackage{graphicx}

\def\rmit#1{{\it #1}}              
\def\specchar#1{{\sc #1}}
\def\FeI{\mbox{Fe\,\specchar{i}}}

\def\SiI{\mbox{Si\,\specchar{i}}}
\def\HeI{\mbox{He\,\specchar{i}}}
\def\CaIIH{\mbox{Ca\,\specchar{ii}\,\,H}}

\def\ie{\rmit{i.e.}}

\def\bibfiles{biblio}
\def\aareferences{\bibliographystyle{iopart-num}
                  \bibliography{\bibfiles}}

\begin{document}
\title{Magneto-acoustic waves in sunspots from observations and numerical simulations}

\author{T Felipe$^{1,2}$, E Khomenko$^{1,2,3}$, M Collados$^{1,2}$ and C Beck$^{1,2}$}

\address{$^{1}$ Instituto de Astrof\'{\i}sica de Canarias, 38205,
C/ V\'{\i}a L{\'a}ctea, s/n, La Laguna, Tenerife, Spain
$^{2}$ Departamento de Astrof\'{\i}sica, Universidad de La Laguna, 38205, La Laguna, Tenerife, Spain
$^{3}$ Main Astronomical Observatory, NAS, 03680, Kyiv,
Ukraine}

\ead{tobias@iac.es}

\begin{abstract}
We study the propagation of waves from the photosphere to the chromosphere of sunspots. From time series of cospatial Ca II H (including its line blends) intensity spectra and polarimetric spectra of Si I $\lambda$ 1082.7 nm and He I $\lambda$ 1083.0 nm we retrieve the line-of-sight velocity at several heights. The analysis of the phase difference and amplification spectra shows standing waves for frequencies below 4 mHz and propagating waves for higher frequencies, and allows us to infer the temperature and height where the lines are formed. Using these observational data, we have constructed a model of sunspot, and we have introduced the velocity measured with the photospheric Si I $\lambda$ 1082.7 nm line as a driver. The numerically propagated wave pattern fits reasonably well with the observed using the lines formed at higher layers, and the simulations reproduce many of the observed features. The observed waves are slow MHD waves propagating longitudinally along field lines.
\end{abstract}

\section{Introduction}

The magnetic field of the sunspots affects the propagation of waves and produces the appearance of new wave modes not present in the quiet Sun. The study of waves in sunspots is interesting from several points of view. It gives information about the atmospheric structure and dynamics. Also, waves are believed to constitute one of the heating mechanisms of the chromosphere. Observations show that the properties of waves greatly vary with the region where they are observed: at the umbral photosphere they present 5 minute period with amplitudes around 100 m s$^{-1}$, while the chromospheric umbral oscillations have basically 3 minute period and amplitudes of a few kilometers per second.

Recent simulations have been able to reproduce many of the observed properties of waves between the photosphere and the chromosphere by means of the introduction of a photospheric pulse which drives a wave spectrum close to the solar one \cite{Felipe+etal2010a}. In this work we go a step further and we have developed numerical simulations where the real photospheric velocity pattern retrieved from observations in the \SiI\ $\lambda$ 1082.7 nm line is used to drive the simulated wave pattern. The oscillations obtained with the simulation at the high photosphere and chromosphere are compared with those measured with lines which are formed at those heights.

\section{Observations}

\begin{figure}[t]
\begin{minipage}{18pc}
\includegraphics[width=18pc]{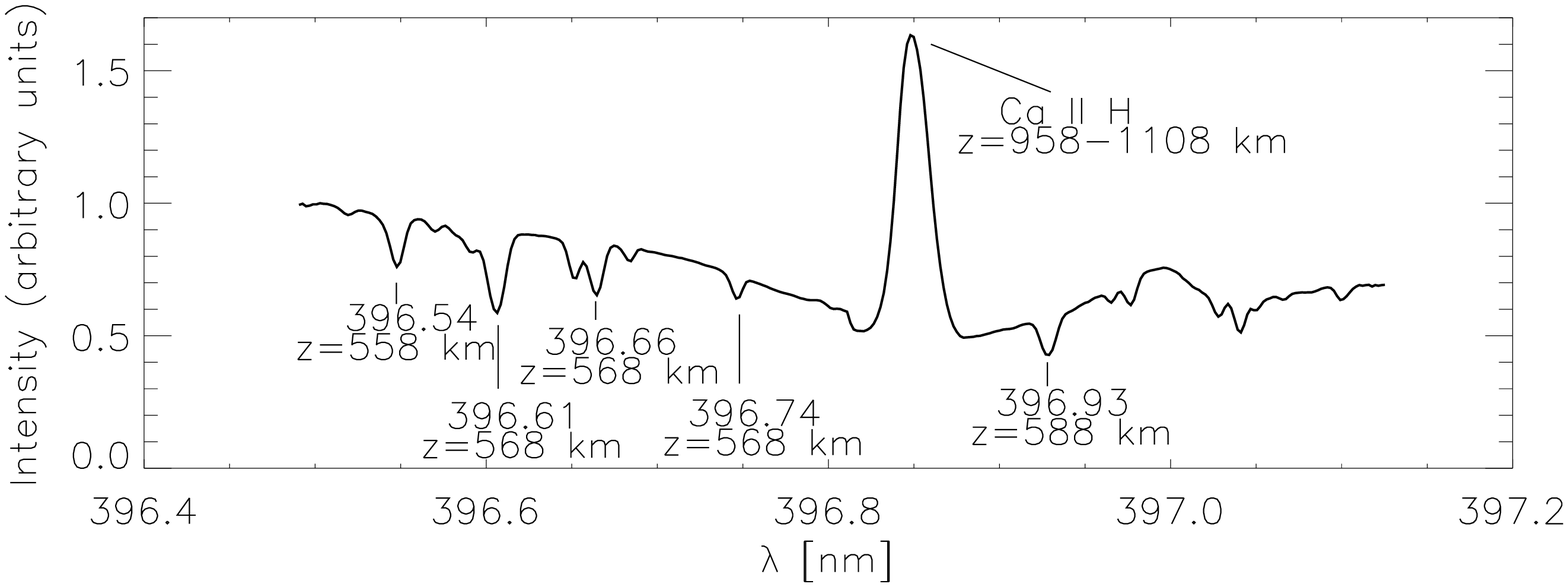}
\caption{\label{fig:spectra_polis}Spectra of the \CaIIH\ line. The lines used and their formation height are marked. }
\end{minipage}\hspace{2pc}%
\begin{minipage}{18pc}
\includegraphics[width=18pc]{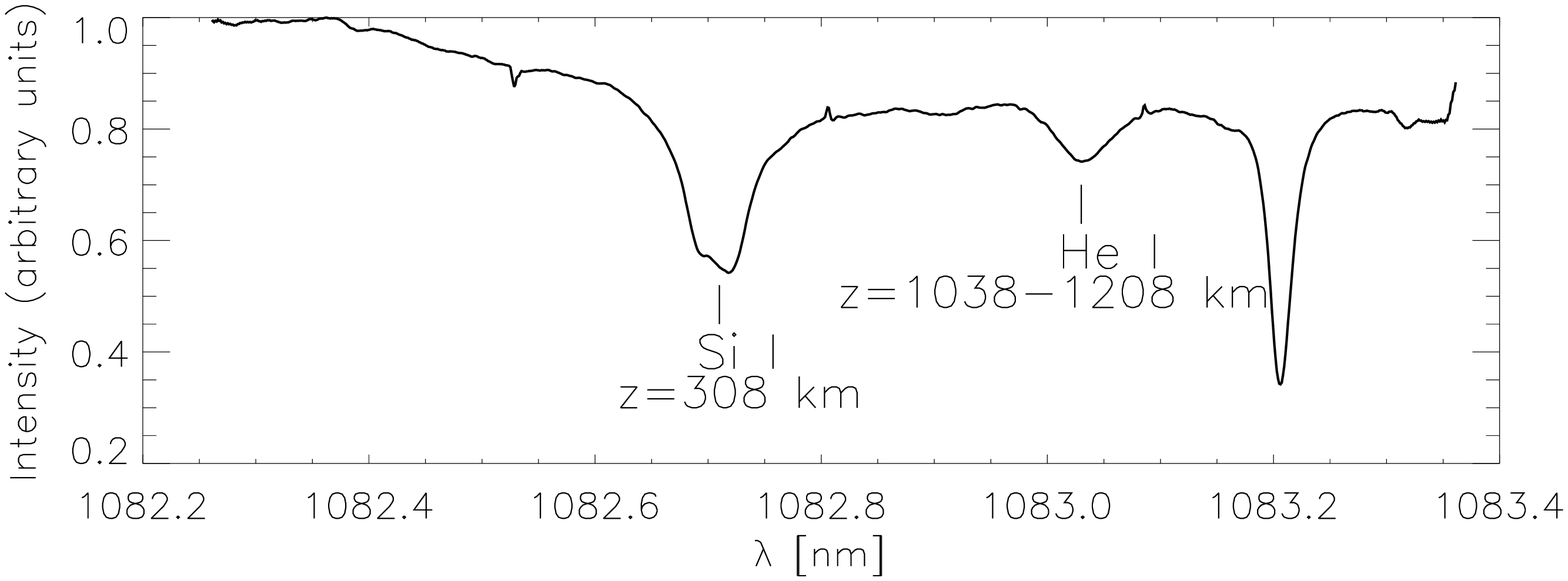}
\caption{\label{fig:spectra_tip}Spectra of the region at 1083 nm. The lines used and their formation height are marked. }
\end{minipage} 
\end{figure}

 We have obtained simultaneous cospatial maps of the LOS velocity measured with photospheric lines (Si I $\lambda$ 1082.7 nm and the Fe I lines at  396.54, 396.61, 396.66, 396.74 and  396.93 nm from the wings of \CaIIH) and chromospheric lines (Ca II H core and He I $\lambda$ 1083.0 nm). The phase and amplification spectra between several pairs of lines in this set of observations have been analized before in \cite{Felipe+etal2010b}. From the fit of the observations to a simple model of linear upward wave propagation we have infered temperature and formation height of the lines. Figures \ref{fig:spectra_polis} and \ref{fig:spectra_tip} show the observed spectral region. The relative formation height of the spectral lines is written next to each line.

\section{Numerical simulations}

The numerical simulations used in this work were performed with a 3D non-linear MHD code \cite{Felipe+etal2010a}. Energy losses are implemented following Newton's cooling law, accounting for the damping of the temperature fluctuations due to radiative losses with Spiegel formula for the typical cooling time \cite{Spiegel1957}. 

We have constructed a MHS model of sunspot using observational data. The stratification of the thermodynamical magnitudes, obtained from the inversion of the umbra and the quiet Sun photospheric spectra of \SiI\ line (using SIR \cite{RuizCobo+delToroIniesta1992}) are set at the axis of the sunspot and at 40 Mm far from it, respectively. These two atmospheres are smoothly merged along the radial distance which separates them \cite{Khomenko+Collados2008}. The  model is an axysymmetric thick flux tube with distributed currents and has no twist. In the horizontal directions the computational domain covers $8.4\times 5.2$ Mm, with an horizontal spatial steps of $\Delta x=\Delta y=100$ km. In the vertical direction it spans form $z=-0.6$ Mm to $z=1$ Mm, excluding the special layer added at the boundaries to absorb the waves without reflections, with a spatial step of $\Delta z=25$ km. The MHS model of the sunspot is pertubed around the photospheric formation height of the \SiI\ line with the velocity measured with this line. This driver generates an oscillatory pattern identical to the observed one at the photosphere, and these waves propagate upward to higher layers.

\section{Results}

In order to compare the numerical simulation with the observational data, we have assigned a fixed $z$ to the formation height of each spectral line, and we assume that the vertical velocity at that height corresponds to the velocity measured from the Doppler shift of the line. (see Figures \ref{fig:spectra_polis} and \ref{fig:spectra_tip}) \cite{Felipe+etal2010b}. 

\begin{figure}[t]
\includegraphics[width=22pc]{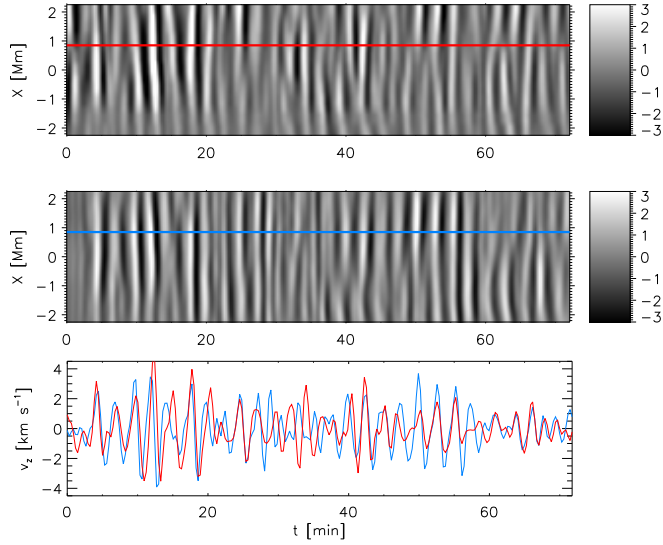}\hspace{1pc}%
\begin{minipage}[b]{14pc}\caption{\label{fig:map}Velocity maps for the chromosphere inside the umbra. Top: observational (measured from the Doppler shift of the \HeI\ line); middle: numerical (vertical velocity at the formation height of the \HeI); bottom: temporal variations of numerical (blue) and observed (red) velocities at $x=0.9$ Mm.}
\end{minipage}
\end{figure}

Figure \ref{fig:map} shows the observational and numerical velocity maps, corresponding to the chromosphere. It is the highest spectral line observed, and these waves have propagated upward about 800 km from the formation height of the \SiI\ line in order to reach this layer. Note that in the simulated velocity map (middle panel) the velocity signal is almost nule during the first 2 minutes, due to the time spent by the slow waves to cover the distance between the driver and this height travelling at the sound speed. During this travel the period of the waves is reduced to around 3 minutes and their amplitude increases, reaching peak-to-peak values of almost \hbox{8 km s$^{-1}$}. Bottom panel of Figure \ref{fig:map} shows that the oscillations develop into shocks. This behavior is well reproduced by the numerical simulation. The simulated velocity map reproduces reasonably well the observed oscillatory pattern. Only in the temporal lapse between $t=27$ and $t=40$ min the simulated pattern differs significatively from the observations.

\begin{figure}[t]
\includegraphics[width=22pc]{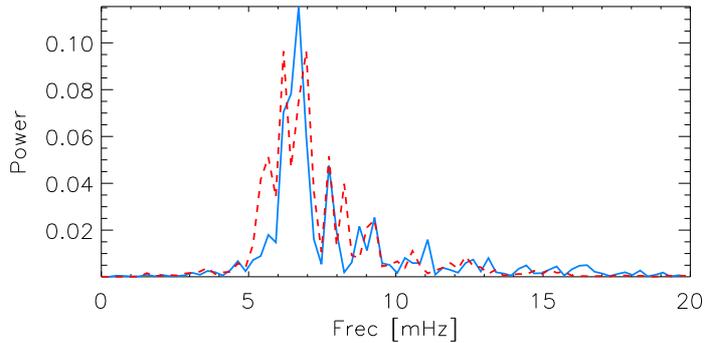}\hspace{1pc}%
\begin{minipage}[b]{14pc}\caption{\label{fig:he_spectra}Power spectra of the velocity measured with the \HeI\ line (red dashed line) and the simulated at the corresponding height (blue line).}
\end{minipage}
\end{figure}

The power spectra at chromospheric heights is dominated by 3 minute oscillations in the band between 5 and 8 mHz (Figure \ref{fig:he_spectra}). The observational power has three power peaks in this band, located at 5.5, 6 and 7 mHz. The simulated power is concentrated at a single peak between the two highest peaks of the observations. The simulated peak at 5.5 mHz is lower than the observed one. The simulations also reproduce the power peaks at 7.7 mHz and 9 mHz, and the low power at frequencies below 5 mHz. At frequencies above 13 mHz the simulated power is slightly higher than the observational one.

\begin{figure}[t]

\includegraphics[width=18pc]{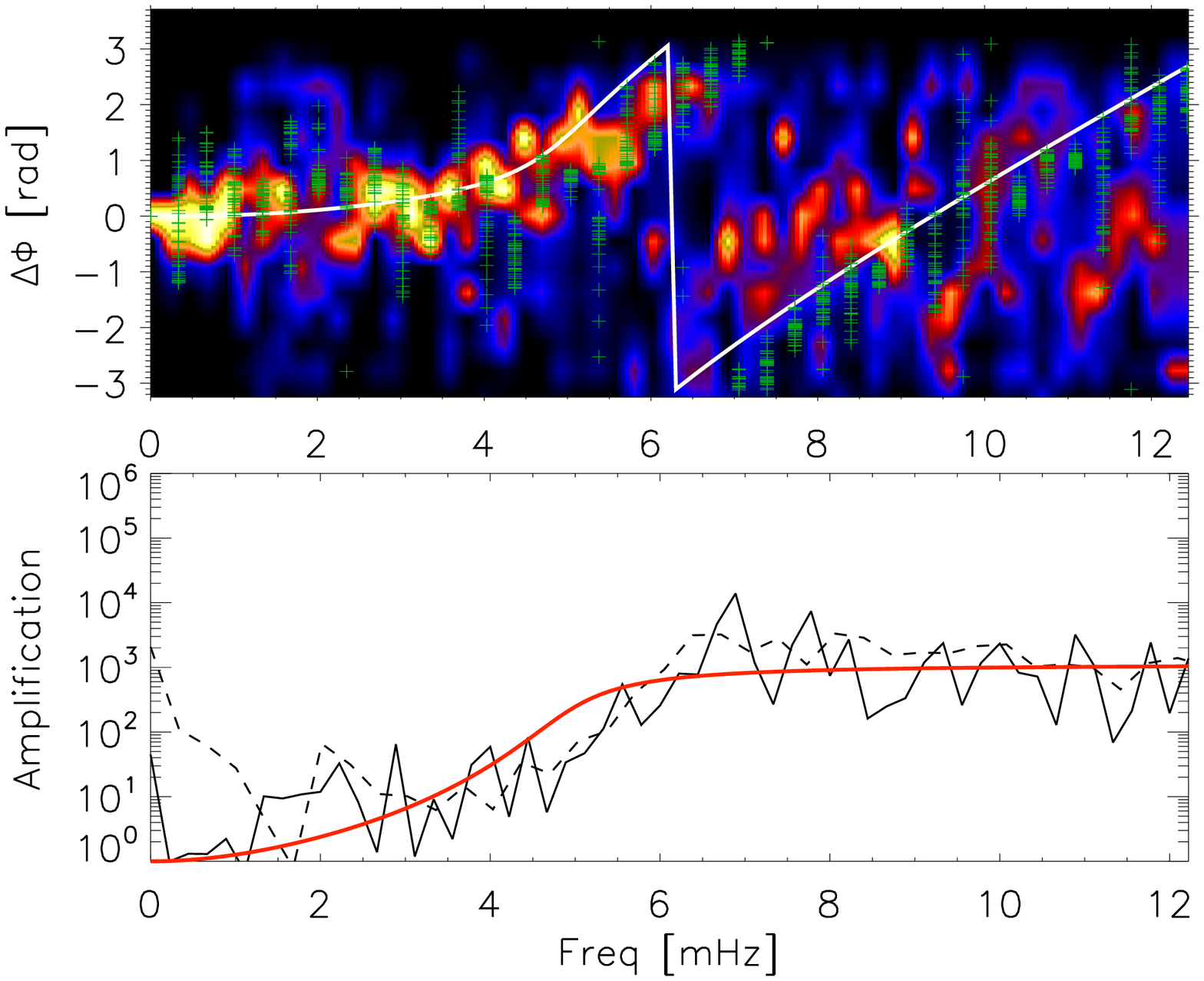}
\includegraphics[width=18pc]{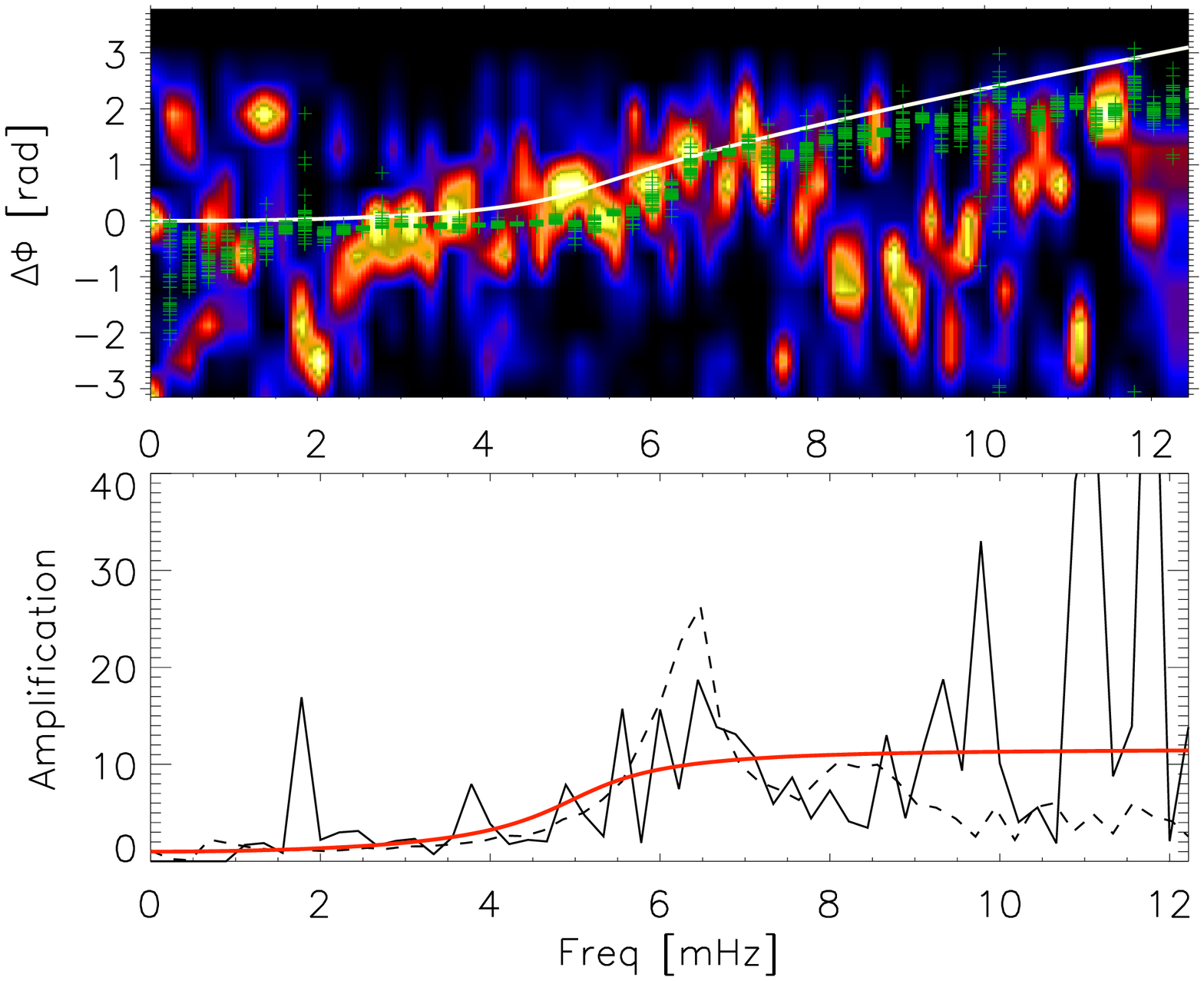}
\caption{\label{fig:dfase} Left panel: Phase difference (top) and amplification spectra (bottom)  between \SiI\ and \HeI\ velocities. Right panel: Phase difference (top) and amplification spectra (bottom)  between \SiI\ and \FeI\ velocities.
The background color in the top panels indicates histograms of the relative occurrence of a given value in the observed phase (from black -low- to yellow -high-). Green crosses are the results of the simulation for all the spatial points. The white line is the fit to a model of linear wave propagation \cite{Felipe+etal2010b}. The bottom panels show the ratio of
the power at the heights of the two signals (amplification spectra) as a function of frequency for the observation (black solid line), the simulation (black dashed line) and the model (red line).
}

\end{figure}

Phase spectra give the delay between two signals at every frequency. The phase difference between the
velocity signals measured in the photospheric \SiI\ line and the chromospheric
\HeI\ line is zero for frequencies below 4
mHz (left panel of Figure \ref{fig:dfase}). At these frequencies the atmosphere oscillates as a whole, \ie, the waves
are stationary. From 4 mHz to 7 mHz, the phase difference increases
linearly with the frequency. It indicates that waves at these
frequencies propagate from the photospheric layer, where the \SiI\
line forms, to the chromospheric layer, where \HeI\ forms. Note the excellent agreement between observational and numerical data from 0 to 7 mHz, where the coherence of the observations is high. At higher frequencies the coherence of the observed phase difference is lower, but the simulated one keeps its linear increase. With regards to the amplification spectra, for frequencies above 1.5 mHz the simulated spectra reproduces properly the observed one.

Right panel of Figure \ref{fig:dfase} shows the phase and amplification spectra between two photospheric lines, the \SiI\ and the \FeI\ $\lambda$ 396.93 lines. The observed phase difference has high coherence between 2 and 8 mHz, and in this frequency range the simulated phase delay fits the observational one reasonably well, showing a $\Delta \phi =0$ for frequencies below 4 mHz and a slow increase for higher frequencies. For frequencies below 2 mHz and above 8 mHz the observed phase difference spreads out and has lower coherence. The behaviour of the simulated amplification spectra is similar to the observed one between 0 and 8 mHz, but the numerical amplification is significatively higher at the peak around 6.5 mHz.

\section{Conclusions}
We have studied the propagation of waves in sunspots from the photosphere to the chromosphere, using observationally driven numerical simulations. We have been able to simulate the travel of the waves through the sunspot atmosphere, with a remarkable match between observations and simulations. The numerical calculations reproduce the wave pattern (Figure \ref{fig:map}), the power spectra (Figure \ref{fig:he_spectra}) and the phase and amplification spectra between several pairs of lines (Figure \ref{fig:dfase}), indicating that observed waves are slow longitudinal acoustic waves dampled by radiative losses.

\section*{References}

\aareferences

\providecommand{\newblock}{}

\end{document}